\documentclass[aps,pre,twocolumn,showpacs,floatfix]{revtex4}
\usepackage{amsmath}
\usepackage{amsfonts}
\usepackage{amssymb}
\usepackage{graphicx}

\begin{document}
\title{ Stochastic energetics of a Brownian motor and refrigerator driven by non-uniform temperature }
\author{Ronald Benjamin}
\email{ronbenin@yahoo.com}
\affiliation{Institut f{\"u}r Theoretische Physik II, Universit{\"a}t D{\"u}sseldorf, D-40225 D{\"u}sseldorf, Germany.}
\date{\today}

\begin{abstract}
The energetics of a Brownian heat engine and heat pump driven by position 
dependent
temperature, known as the B{\"u}ttiker-Landauer heat engine and heat pump, is
investigated by numerical simulations of the inertial
Langevin equation. We identify  parameter values for optimal performance of the 
heat engine
and heat pump. Our results qualitatively differ from
approaches based on the overdamped model.
The behavior of the heat engine and heat pump,
in the linear response regime is examined under finite time conditions
and we find that the efficiency  is lower than that of an endoreversible
engine working under the same condition.
Finally, we investigate the role of
different potential and temperature profiles to enhance the efficiency
of the system. Our simulations show that optimizing the potential
and temperature profile leads only to a marginal enhancement
of the system performance due to the large entropy production via the Brownian
particle's kinetic energy.
\end{abstract}


\maketitle

\section{ Introduction}
\label{sec:introdution}

It has been  known since the works of Landauer~\cite{landauer88}, Van 
Kampen~\cite{vankampen88,vankampen91} and
B{\"u}ttiker~\cite{buttiker87}  that a Brownian particle in a periodic potential 
and subject to 
a spatially inhomogeneous and periodic temperature profile can move 
preferentially
in one direction.  In presence of a small external load it can also
do work as a heat engine. Furthermore,
according to Onsager symmetry the motor can also work as a refrigerator or heat 
pump, transferring
heat from the cold to the hot bath in presence of an external power supply.
This mechanism of directed transport and heat transport in presence of spatially 
non-uniform heat bath,
is well known as the B{\"u}ttiker-Landauer heat engine and heat pump.
Along with the Feynman-Smoluchowsky ratchet and pawl heat 
engine~\cite{feynman,reimann02}, this was one of the
first examples of  autonomous Brownian motors, which have subsequently been 
widely studied 
by many authors in recent 
years~\cite{bierastu96,derenyi99,hondou00,matsuo00,asfaw04,ai05+06,goko2005,
benjamin08}.

The BL heat engine/heat pump is an ideal system to investigate
non-equilibrium thermodynamics of small scale systems where thermal fluctuations 
become 
important~\cite{parrondo02}. 
Thermodynamic quantities such as the work and heat
which determine how effectively such microscopic engines perform their tasks
and are also of great interest to the nanoscale industry desirous of  building 
artificial Brownian motors. 
Sekimoto in 1997 formulated stochastic energetics~\cite{sekimoto97},
based on the Langevin equation,
to quantify the heat flow and efficiency of such Brownian engines.

Several authors have investigated the energetics of the B{\"u}ttiker-Landauer 
heat engine
and heat pump using Sekimoto's formulation.
It was shown using overdamped models (mass of the Brownian 
particle $M=0$)~\cite{asfaw04,asfaw-bekele2007,matsuo00} based on the 
Fokker-Planck equation 
as well as based on other phenomenological approaches~\cite{ai05+06},
that the BL heat engine can reach Carnot efficiency and the BL heat pump can 
attain the
corresponding Carnot coefficient of performance. However,
Der{\`e}nyi-Astumian~\cite{derenyi99} and Hondou-Sekimoto~\cite{hondou00} later
argued that Carnot efficiency is unattainable due to the irreversible heat
flow via kinetic energy from the hot to the cold bath whenever the Brownian 
particle 
crosses a temperature boundary. 
They made phenomenological predictions regarding the failure
of overdamped models in predicting this heat transfer and argued that the
heat flow via kinetic energy diverges in the overdamped limit ($M\rightarrow 
0$). 

In a recent work~\cite{benjamin08}, stochastic energetics
of the B{\"u}ttiker-Landauer heat engine and heat pump was investigated
by a direct comparison of numerical solution of the inertial Langevin equation 
with molecular dynamics simulations.  Good agreement between the two approaches
confirmed Sekimoto's stochastic energetics as well as the predictions of
Der{\`e}nyi-Astumian~\cite{derenyi99} and Hondou-Sekimoto~\cite{hondou00} and 
showed
that the overdamped model ($M=0$) is not equivalent to the overdamped limit 
($M\rightarrow 0$).
In fact one should take into account the inertial mass of the Brownian particle 
and then go to
the overdamped limit. At the quasistatic limit corresponding to zero average 
motor velocity,
where overdamped models predict Carnot efficiency, the true efficiency is 
actually zero because
even if heat flow via potential vanishes, the irreversible heat flow via kinetic 
energy does not 
vanish. This irreversible heat also dominates the heat flow via the potential 
and greatly reduces 
the efficiency of the heat engine and the coefficient of performance of the heat 
pump.

Carnot efficiency, if attainable, is reached only in the quasistatic limit, when 
the 
operating time of the engine is infinite and the power output is thus zero. 
Since a real heat engine must 
produce power in a finite time, it is more appropriate to
analyse its efficiency under the condition of maximum power.
In a recent work, Van den Broeck~\cite{vandenbroeck05} has shown that in
the linear response regime, the efficiency at maximum power of a Brownian engine
reaches the efficiency of an endoreversible engine when it operates at
maximum power, namely the Curzon-Ahlborn efficiency~\cite{curzon75}. 
Corresponding to the efficiency at
maximum power, some authors have studied the equivalent quantity for the heat 
pump using linear
irreversible thermodynamics~\cite{cisneros06}.  Analytical expressions for the 
efficiency at 
maximum power~\cite{vandenbroeck05,izuoku2010,izuoku2012,Tu08} and the 
corresponding quantity for
a heat pump have been derived but the validity of these expressions have 
not been confirmed for the BL heat engine/heat pump for a finite mass of the 
Brownian particle.

The efficiency at maximum power
of the BL heat engine has  been studied using overdamped models~\cite{asfaw04}, 
and
the authors showed that it is close to the Curzon-Ahlborn efficiency.
Gomez-Marin and Sancho~\cite{alexsanchoPRE06} studied a certain model of the BL 
heat engine/heat pump
based on the overdamped Fokker-Planck equation and concluded
that efficiency at maximum power in the linear response regime is equal to 
Curzon-Ahlborn efficiency. 
However, as was pointed out by the them, a proper analysis of the thermodynamics
of the BL heat engine in the linear response regime can only be made by 
considering the inertial term since 
irreversible heat flow via kinetic energy plays an important role in its 
energetics.

While the dominant contribution to the heat flow comes from the kinetic energy 
contribution,
many authors ignore the kinetic energy contribution and discuss ways to enhance 
efficiency of 
the BL heat engine based on an overdamped approach. Some papers have reported a 
very high efficiency for the 
BL heat engine, of the same order as the Carnot efficiency,  in certain 
parameter regimes.
These works discuss various strategies to enhance the efficiency of the BL heat 
engine by modifying
the potential and temperature profiles. However, such approaches towards 
designing a more efficient  heat engine
have not been tested by considering mass of the Brownian particle and thus the 
heat flow via kinetic energy.

Since, kinetic energy plays such a crucial role in presence of spatially 
inhomogeneous temperature, a more 
complete description of the thermodynamics of the BL heat engine requires taking 
into consideration mass of the Brownian 
particle explicitly. As kinetic energy contribution dominates over the heat 
transfer via the potential,
one should first attempt to reduce the kinetic energy contribution and then 
discuss other ways to enhance the 
efficiency. While some works have tried to include the irreversible heat flow 
via kinetic 
energy in a phenomenological 
manner in their theoretical 
calculations~\cite{asfawepjb2008,zhang2011,linchen09}, a direct 
quantification of this irreversible heat
is not to be found in the literature apart from the simulation study of 
Ref.~\cite{benjamin08}. However,
in Ref.~\cite{benjamin08}, only one model of the BL heat engine was investigated 
and there was no discussion on 
possible ways to 
enhance the efficiency of the heat engine. Moreover, energetics of the Brownian 
heat pump or refrigerator based on 
a spatially non-uniform temperature was not considered.

In the present work, we attempt to fill an important gap in the study on the 
thermodynamics
of the BL heat engine by explicitly taking into account the mass of the Brownian 
particle.
As we show in the next sections,
our results qualitatively differ with previous studies on the efficiency 
optimization
of the BL heat engine based on overdamped approaches. 
We will investigate the energetics of the BL heat engine and heat pump by 
extensive numerical simulations 
of the Langevin equation. First, we numerically explore the optimal conditions 
for  performance of the 
BL engine. Then we will discuss the operation of the system
in the linear response regime and test the validity of the expressions for the 
efficiency and coefficient of power 
under finite time conditions. Finally we consider different temperature and 
potential profiles to diminish the kinetic 
energy contribution and enhance the performance of the heat engine and heat 
pump.

The paper has been organized as follows:
in the next section we introduce the basic model, as studied in a
previous work~\cite{benjamin08}, and the methods used to investigate
the  energetics of the BL heat engine and heat pump.
In sections~\ref{sect:eff_motor} and \ref{sect:perf_ref}, we discuss the
efficiency of the BL heat engine and heat pump respectively based on the
basic model. Section~\ref{sect:linear} discusses the efficiency and
coefficient of performance in the linear response regime.
In  section~\ref{sect:effopt}, we study the effect of different
potential and temperature profiles on the heat engine efficiency and
heat pump coefficient of performance. Finally, we end with a conclusion.

Throughout this paper, we use the terms motor and heat engine interchangeable to
mean the same thing i.e. a Brownian motor doing work against an external load.
Similarly the terms heat pump and refrigerator mean the same thing.

\section{Basic model and methods}
\label{sect:basic_model}

\begin{figure}
\includegraphics[width=3.0in]{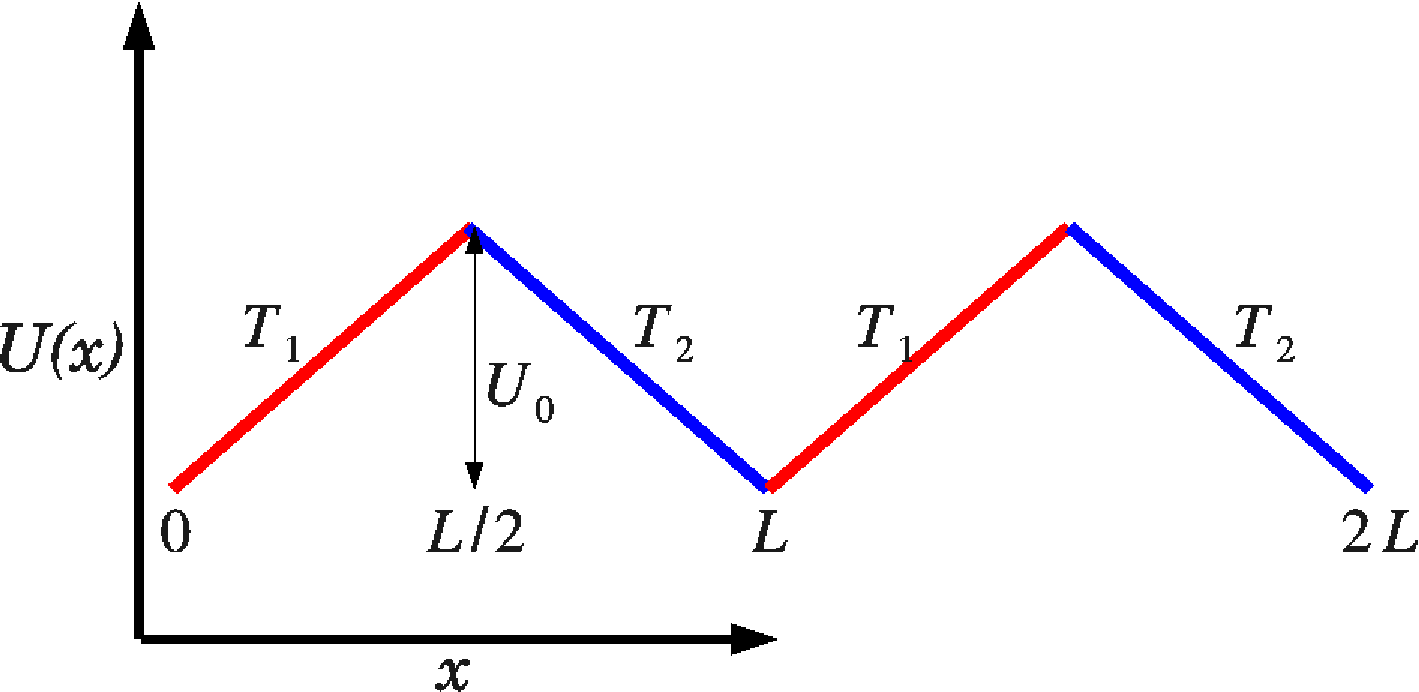}
\caption{\label{fig:model}(Color online) Two rectangular reservoirs
at temperatures $T_{1}$ and $T_{2}$ are
alternately connected. Brownian
particles are subjected to a piece-wise linear potential.}
\end{figure}

We consider a chain of two-dimensional cells attached
to each other along the $x$ direction (see Fig.~\ref{fig:model}).
Each cell is $L/2$ long and filled with  heat bath particles.
The cells are thermally isolated
from each other and at  thermal equilibrium with a temperature independent
from the neighbouring cells. Brownian particles of mass $M$ are placed
in the cells and, unlike the heat bath
particles, can move through the cell walls. They are also
subjected to a periodic piece-wise linear potential $U(x)$
with a period $L$ in the $x$ direction:
\begin{equation}
U(x) =
\begin{cases}
-\frac{2 U_{0}}{L} x & \text{for  $ -\frac{ L}{ 2} < x
\leq  0$}, \\
\frac{2 U_{0}}{L}x & \text{for $  0 < x \leq  \frac{L}{2} $},
\end{cases}
\label{eq:U}
\end{equation}
where $U_{0}$ is the potential height. In addition to the
periodic potential,  a constant
external force $F$ is exerted on the Brownian particles.  Instead of infinitely
long chains we consider only two cells with a periodic boundary condition: cell 1
with temperature $T_{1}$ for $-L/2 < x \leq 0$ and cell 2 with $T_{2}$ for $0 <
x \leq L/2$.

The motion of the Brownian particle in the $x$ direction can be
investigated by the one-dimensional Langevin equation:
\begin{equation}
\begin{gathered}
\dot{x} = v, \\
M\dot{v}=-\gamma(x) v- U^\prime(x) + F + \sqrt{2 \gamma(x) T(x)}
\xi(t),
\end{gathered}
\label{eq:Langevin}
\end{equation}
where $x$ and $v$ are the position and velocity of the Brownian particle and
$\xi(t)$ is a standard Gaussian white noise:
\begin{equation}
\langle \xi(t)\rangle=0, \quad \langle \xi(t)\xi(s)\rangle
=\delta (t-s).
\end{equation}

The piecewise constant temperature is given by:
\begin{equation}
T(x) =
\begin{cases}
T_{1} & \text{for  $ -\frac{\displaystyle L}{\displaystyle 2} < x \leq 0 $},\\
T_{2} & \text{for  $ 0 < x \leq \frac{\displaystyle L}{\displaystyle 2} $}.
\end{cases}
\label{eq:T}
\end{equation}
Temperature is measured in energy unit ($k_{B}=1$).
Here, and later on, an overdot refers to derivative taken with respect to time
and a prime means derivative taken with respect to space. Assuming that
Einstein's
relation, namely, $D(x)=T(x)/\gamma(x)$ holds locally, the friction coefficient
also depends on the position in the same way as the temperature through the
relation
\begin{equation}
\gamma(x)=\rho\sigma_{B}\sqrt{2\pi T(x)m},
\end{equation}
where, $\rho$, $m$ and $\sigma_{B}$ refer to the number density of heat bath 
particles in a cell 
($\rho=0.01$ in all our simulations), the mass of a bath particle and the 
diameter of the Brownian particle, respectively.
The above analytical expression for $\gamma(x)$ holds for an ideal
gas~\cite{meurs04}. Since, in our previous work~\cite{benjamin08}, Langevin 
equation results
were compared with hard-disk simulations of a Knudsen gas, we use the same
dependence of the friction coefficient on the temperature for our Langevin 
simulation. 
The length of the simulation box was $L=500$. We  used the Heun 
algorithm~\cite{kloeden} to numerically 
solve Eq.~\ref{eq:Langevin} with a time-step  of $\tau=0.001$ ($M/\gamma > 1.0$ corresponding
to the underdamped regime) or $0.0001$ ($M/\gamma \leq1.0$ corresponding to the overdamped regime). 
For ease of numerical implementation all of our simulations were done in 
reduced units~\cite{reimann02}.

In order to investigate the thermodynamic properties, we use stochastic
energetics formulated by Sekimoto~\cite{sekimoto97,sekimoto98,sekimoto}. The
heat flux from the gas particles in the $i$-th cell to the Brownian particles is
defined as,
\begin{equation}
 \dot{Q}_i = \left \langle \left (-\gamma_i \dot{x} + \sqrt{2 \gamma_i
T_i}
\xi(t) \right )  \dot{x} \right \rangle_i\;,
\label{eq:heat_definition}
\end{equation}
where, $\langle \cdots \rangle_i$ indicates ensemble average taken while the
Brownian particles are located in the $i$-th cell. Using the  Langevin
equation~(\ref{eq:Langevin}), we obtain the total heat flux as a sum of three
terms:
\begin{align}
 \dot{Q}_i &=  \frac{M}{2} \frac{d}{dt}\left \langle \dot{x}^2 \right
\rangle_i
+ \left \langle U^\prime (x)  \dot{x} \right \rangle_i
- F \left \langle \dot{x} \right \rangle_i \notag \\
&=\dot{Q}^\text{\sc KE}_i + \dot{Q}^\text{\sc PE}_i + \dot{Q}^\text{\sc
J}_i
\label{eq:heat_inertial_Langevin}
\end{align}
where the first two terms on the rhs are the kinetic energy and potential energy
contribution to the heat flux, respectively, and the last term is the Joule
heat. Rate of work done on a Brownian particle (or the power), by the external force $F$ in each cell
is given by,
\begin{equation}
\dot{W}_{i}=F\langle \dot{x} \rangle_{i}.
\end{equation}
and the total power is given by,
\begin{equation}
\label{eq:work}
\dot{W}=F(\langle \dot{x}\rangle_{1}+\langle \dot{x}
\rangle_{2})=F\langle\dot{x}\rangle
\end{equation}
where, $\langle \dot{x}\rangle_{1}=\langle \dot{x}\rangle_{2}=\langle
\dot{x}\rangle /2$,
since width of both cells are the same and, $\langle \dot{x} \rangle =\langle v \rangle$
is the net particle current. Since, $\dot{W}$ represents the total rate
of work done on the Brownian particle, the power output in case of a heat engine is
$-\dot{W}$. When the system functions as a heat engine in presence of an external load ($F<0$),
the net velocity $v>0$ and therefore the power output $-\dot{W}>0$.
In case of a heat pump, $\dot{W}$ represents the supplied power or the power input.

In the steady state, the net energy flux to the Brownian particles will vanish,
and thus the energy gained by the Brownian particles in cell 1
will be cancelled by the energy loss in cell 2.  Therefore, heat flux
from cell 1 to cell 2 via the Brownian particles is defined as
\begin{align}
 \dot{Q}_{1 \rightarrow 2} &= \dot{Q}_1 + \dot{W}_1 =  \dot{Q}^\text{\sc KE}_1 +
\dot{Q}^\text{\sc PE}_1\, \notag \\
 &= -\dot{Q}_2-\dot{W}_2=-\dot{Q}^\text{\sc KE}_2 -
\dot{Q}^\text{\sc PE}_2\;.
\label{eq:heat_transport}
\end{align}

The above equation represents the first law of thermodynamics as applied to the
BL heat engine, which can also be written as,
\begin{equation}
 \dot{Q}_{1}=-\dot{Q}_{2}-\dot{W}
\end{equation}
which tells us that part of the net heat extracted from the hot bath is used
by the Brownian particle to perform work against the external load and the rest is
dissipated as heat to the cold bath.

A Brownian heat engine is in a non-equilibrium state and  according
to the second law of thermodynamics, the net entropy production during
the working of the motor must be positive. For the BL motor described above, 
which operates under the influence of two heat baths at different temperatures, the
entropy production according to the second law is~\cite{parrondo02},
\begin{equation}
 \dot{S}=-\frac{\dot{Q}_{1}}{T_1}-\frac{\dot{Q}_{2}}{T_{2}}\geq 0
\end{equation}

The thermodynamic efficiency of the BL
heat engine, is defined as
\begin{equation}
\eta=\frac{-\dot{W}}{\dot{Q}_{1}}\;.
\end{equation}
and using the first and second laws, can be written as,
\begin{equation}
\eta=\eta_{\rm c}-\frac{\dot{S}}{\dot{Q}_{\rm 1}}T_{1} 
\end{equation}
where, $\eta_{\rm c}=1-T_2/T_1$ is the Carnot efficiency.

Similar to the efficiency of the heat engine, one can define the coefficient of
performance as,
$\varepsilon$, defined as,
\begin{equation}
\label{eq:cop}
 \varepsilon=\frac{\dot{Q}_{2}}{\dot{W}},
\end{equation}
and using the first and second laws of thermodynamics, Eq.~\ref{eq:cop}
can we written as,
\begin{equation}
\varepsilon=\frac{ \varepsilon_{\rm c}  \dot{Q}_{\rm 2} } {  \varepsilon_{\rm c} T_{\rm 1}\dot{S} +\dot{Q}_{\rm 2}  }
\end{equation}
where, $\varepsilon_{c}=T_{2}/(T_{1}-T_{2})$.

The respective Carnot limits for the heat engine efficiency and heat pump
coefficient of performance are attainable only if the entropy production $\dot{S}$ vanishes,
such that the engine works in a reversible way. However, Carnot efficiency,
if attainable, is possible only under a quasistatic condition when the
engine does no work or the heat pump transfer no heat. For the BL heat 
engine/heat pump however, there is production of entropy
even in the quasistatic limit and hence it never attains the 
Carnot efficiency~\cite{benjamin08}.
On the other hand, any finite power output or heat transfer will always be accompanied
by entropy production which limits the efficiency/coefficient of performance of the 
heat engine/heat pump and part of the effort in this
paper is devoted to finding out possible models with minimal
entropy production leading to a higher efficiency, though without
much success, as we will find out in the latter sections.

\section{Efficiency of the Heat Engine}

\label{sect:eff_motor}

\begin{figure}
\begin{center}
\leavevmode
\includegraphics[width=3.0in]{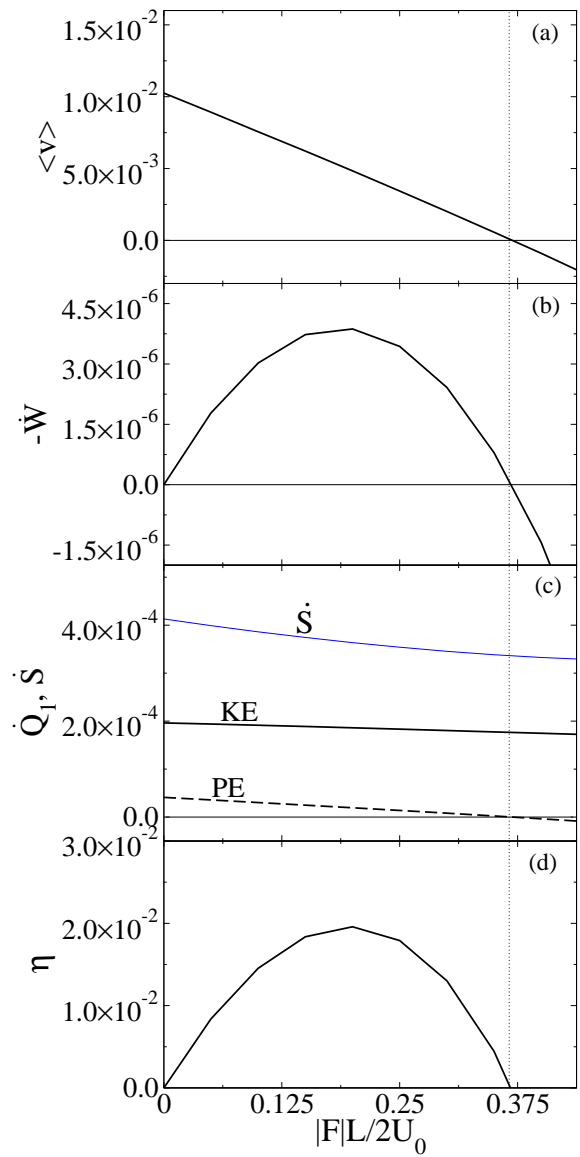}
\caption{\label{fig:motor_qke}  (a) Average velocity $\langle v \rangle$,
(b) power $-\dot{W}$, (c) $\dot{Q}_{1}^{\text{KE}}$,
$\dot{Q}_{1}^{\text{PE}}$ and $\dot{S}$,
and (d) efficiency of the heat engine as a
function of $|F|L/2U_{0}$.  In (a), (b) and, (c), the thin horizontal
solid line represents $y=0$.  The vertical line coincides with the stall load
$F_{\rm stall}$, where the motor velocity, work output and efficiency go
to zero.
The parameter values are $T_{1}=0.7$, $T_2=0.3$,
$\sigma_{B}=5$ and $M/m=5$.  
The Carnot efficiency $\eta_{c}=0.57$.}
\end{center}
\end{figure}

When two cells have different temperatures ($T_{1}>T_{2}$),
Brownian particles in the hotter cell can reach
the higher-potential energy region more often
 than those in the low-temperature cell.
As a result, the Brownian particles tend to move
from the hot to the cold cell over the potential barrier, resulting
in a net current in the positive direction even  in the absence of
an external force ($F=0$). In the presence of an external load ($F<0$),
the Brownian particles can do work against it behaving as a motor.

In Fig.~\ref{fig:motor_qke}, we plot the velocity,  
the power ($-\dot{W}$), various components of the heat
flow out of cell $1$ at temperature $T_{\rm 1}$, the 
entropy production $\dot{S}$ and efficiency as a
function of the external load. As the external load impedes
the motion of the Brownian particles, the velocity decreases and
vanishes at the stall load $F_{\rm stall}$. When $F=0$, the average
velocity of the Brownian particles is not zero, but the
work is zero. At the stall load, since the motor velocity
goes to zero, the power output again vanishes. At a certain
value of $F$, between $0$ and $F_{\rm stall}$, close to $F_{\rm stall}/2$,
the power output reaches a maximum i.e., the Brownian heat engine
 delivers maximum power.
Now we come to the heat transfer. When the Brownian
particle crosses one period to the right it extracts
potential energy $U_{0}$ from the higher-temperature cell at
the rate $(2U_{0}/L)\langle \dot{x}\rangle_{1}$ and dissipates the same amount of
potential energy to the cold cell. This heat flow via potential
decreases with the load as the velocity decreases
and ultimately vanishes at $F_{\rm stall}$. At the quasistatic limit
of zero velocity, this heat transfer via potential energy
is reversible.

On the other hand, Fig.~\ref{fig:motor_qke} shows that the heat flow via
kinetic energy varies slowly with the load and is positive
even in the quasistatic limit, showing that a net heat
flows from the hot to the cold reservoir even when the
heat engine does not deliver any work. As explained in a
previous work~\cite{benjamin08}, when the Brownian particle crosses a
temperature boundary  its kinetic energy thermalizes to the 
new temperature thereby dissipating heat. 
Even when the motor velocity is zero at the stall load $F_{\rm stall}$,
the Brownian particles are still crossing the temperature boundary back
and forth. Therefore, this heat continues to flow at the
stall load. Since the heat flows always from the hot cell
to the cold cell, independent of the direction of the Brownian
particle's movement, it is irreversible heat. As Fig.~\ref{fig:motor_qke} shows,
at the stall load, the heat flow via potential is zero but the kinetic energy contribution
is still positive leading to a non-zero entropy production.

The presence of this entropy producing irreversible heat carried by the kinetic
energy of the Brownian particle, makes it impossible for the BL heat engine to reach Carnot
efficiency, $\eta_{c}=1-T_{1}/T_{2}$.  In fact, Fig.~\ref{fig:motor_qke} shows that
the efficiency is an order of magnitude lower than $\eta_{\rm c}$ and diminishes to zero
at the quasistatic limit. 

Beyond the stall load, the motor reverses its velocity and moves in the direction
of the external load. In this situation work is done on the Brownian particle
and hence the power output is negative and so is the heat flow via potential.

Figure~\ref{fig:motor_delt} shows the power,
heat flow via kinetic energy,
and heat engine efficiency as a function of the temperature difference
between the two cells. Overdamped models~\cite{asfaw04}, which
ignore the kinetic energy contribution, predict
that the efficiency increases with $\Delta T$ and then saturates. However,
when we take inertia into account, its a different story.
At smaller $\Delta T$, the power output is low leading to
a low efficiency while at large $\Delta T$, the heat flow via kinetic energy
is large, again reducing the efficiency. While the power output
saturates with increasing temperature difference between the cells, the entropy production 
and heat flow via kinetic energy 
monotonically increase with $\Delta T$.
There is an optimum $\Delta T$ at which
the efficiency is maximum, but it is far lower than the Carnot efficiency. 
We also carried out simulations where the friction coefficients in both
the hot and cold baths were taken to be the same, but observed the same dependence
of the thermodynamic quantities on the temperature difference between the cells.

It was shown in a previous work~\cite{benjamin08} that $\dot{Q}^{KE}$
diverges as $M^{-1/2}$ in the overdamped limit $(M \rightarrow 0)$.
This can be understood as follows: the Brownian particle
thermalizes over a length scale $l_{\rm th}= v_{\rm th}\tau =\sqrt{TM}/\gamma$,
where $v_{\rm th} =\sqrt{T/M}$ is the thermal velocity and $\tau=M/\gamma$
is the typical relaxation time. Suppose a Brownian particle
moves from a hot ($T_{1}$) to a cold ($T_{2}$) cell. When the particle crosses the border,
it dissipates an amount of heat $(1/2)(T_{1}-T_{2})$ to the cold cell and then moves onward to the
next hot cell or recrosses back to the same hot cell. As the overdamped limit ($M\rightarrow 0$)
is approached, the thermal length scale goes to zero and the Brownian 
particle thermalizes more quickly to its new environment as it 
crosses a temperature boundary and  thereby dissipates
more heat in the same time period. The effective temperature gradient over
which this thermalization of the kinetic energy occurs is $|T_{1}-T_{2}|/l_{\rm th} \propto M^{-1/2}$
and as $M\rightarrow 0$, this kinetic energy contribution diverges. Consequently,
the entropy production also goes to infinity in the overdamped limit ($M\rightarrow 0$).
Since this irreversible
heat dominates the heat flow via potential, the
efficiency diminishes to zero as $\sqrt{M}$, in the overdamped
limit.

\begin{figure}
\begin{center}
\leavevmode
\includegraphics[width=3.0in]{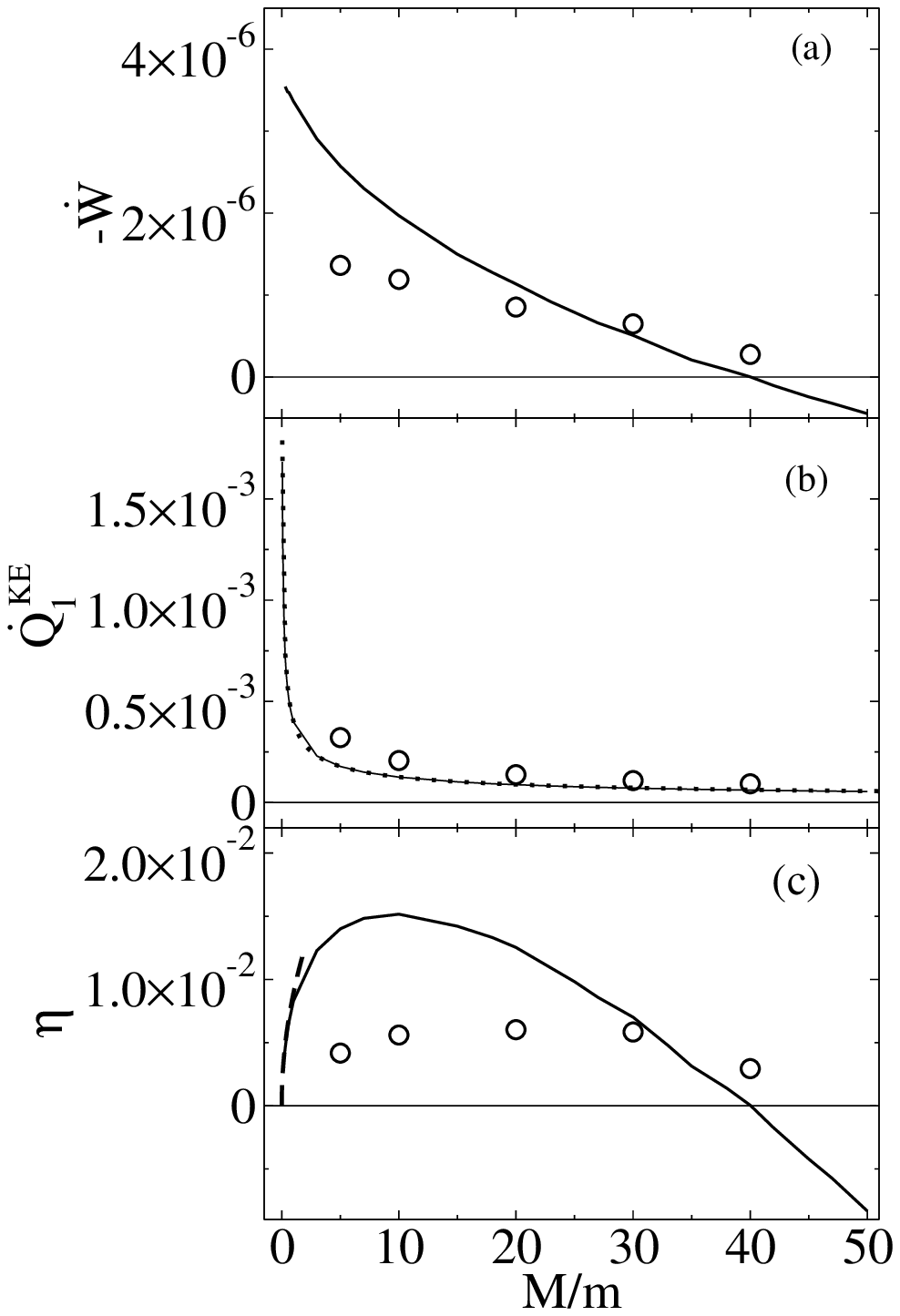}
\caption{\label{fig:motor_mass} (a) Power output, (b) heat flux due to kinetic
energy (solid line) and entropy production $\dot{S}$ (dashed line)  and, 
(c) efficiency as a function of the mass ratio $M/m$.
 The dotted line in (b) corresponds to a phenomenological fit $(\propto M^{-1/2})$
to the kinetic energy contribution. The dotted line in (c) represents
a fit ($\propto \sqrt{M}$) to the efficiency in the overdamped limit
($M\rightarrow 0$). In (a), the thin horizontal
solid line represents $y=0$.
The parameter values are $T_{1}=0.7$,  $T_2=0.3$, $|F|L/2U_{0}=0.3$ and
$\sigma_{B}=5.0$. The Carnot efficiency $\eta_{c}=0.57$.}
\end{center}
\end{figure}

On the other hand, in the underdamped limit $(M\rightarrow \infty)$, 
the Brownian particle doesn't thermalize to the local
environment before entering the next heat bath~\cite{blanter98} and
the motor fails, diminishing the power and efficiency to
zero. The efficiency will be maximum at an optimal value
of $M$. Figure~\ref{fig:motor_mass}, clearly confirms these predictions. 
Simulations were also carried out for same friction coefficients in both
the hot and cold baths, but the qualitative dependence on the mass
of the Brownian particle remains the same.

\begin{figure}
\begin{center}
\leavevmode
\includegraphics[width=3.0in]{fig3.eps}
\caption{\label{fig:motor_delt} (a) Power output, (b) heat flux due to kinetic
energy and the entropy production ($\dot{S}$) and (c) 
efficiency as a function of  ${\Delta T} = T_{1}-T_{2}$,
where $T_{2}$ is kept at $0.1$. 
The parameter values are
$M/m=5.0$, $|F|L/2U_{0}=0.5$ and $\sigma_{B}=5.0$. In (a), the thin
horizontal
dashed line represents $y=0$.
The Carnot efficiency
varies from $0.67$ at $\Delta T=0.2$ to $0.96$ at $\Delta T=2.8$.}
\end{center}
\end{figure}

%

\begin{figure}
\begin{center}
\leavevmode
\includegraphics[width=3.0in]{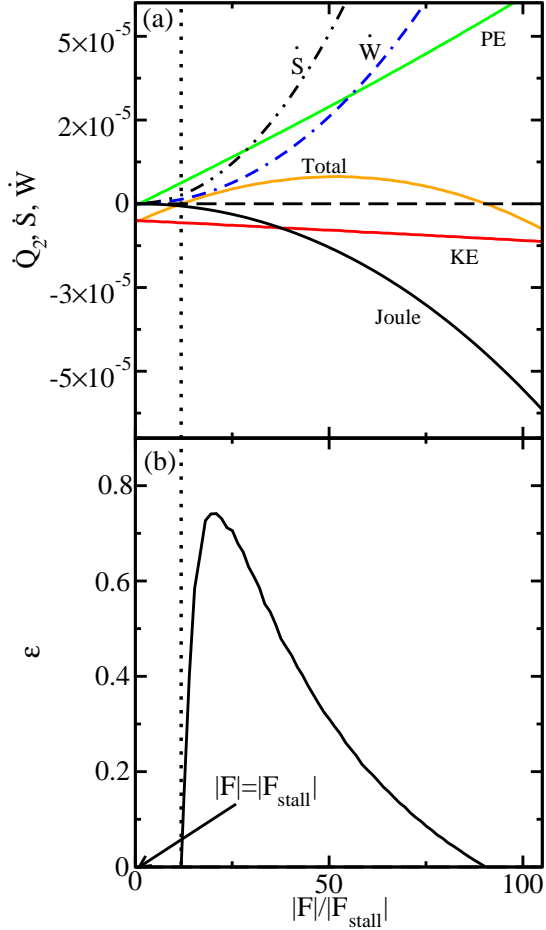}
\caption{\label{fig:fridge_nonuniform} (a)Power input $\dot{W}$ (blue double dash-dotted line),
various components of $\dot{Q}_{2}$ [from top to
bottom the heat flows are $\dot{Q}_{2}^{\text {PE}}$ (green line),
$\dot{Q}_{2}$ (orange line), $\dot{Q}_{2}^{\text {KE}}$ (red line) and 
$\dot{Q}_{2}^{\text{J}}$ (black line)], the entropy production $\dot{S}$ (black dash double dotted line) 
and (b) the coefficient
of performance $\varepsilon$ as a function of $F/F_{\rm stall}$,
$F_{\rm stall}$ being the stall load. In (a), the thin horizontal
dashed line represents $y=0$. The vertical line coincides with the value
for the load where the cooling begins i.e. 
$|\dot{Q}_{2}^{\text {PE}}|>|\dot{Q}_{2}^{\text {KE}}|+|\dot{Q}_{2}^{\text{J}}|$.
The parameter values
are ${\Delta T}=T_{1}-T_{2}=0.01$, $T_{2}=0.5$, $M/m=5.0$ and $\sigma_{B}=5.0$.
The Carnot coefficient of performance $\varepsilon_{\rm c}=50$.}
\end{center}
\end{figure}

\section{Performance of Heat Pump}
\label{sect:perf_ref}

We now discuss the energetics of the BL heat pump.
When the external load exceeds the stall load, Brownian particles
reverse their direction to move
in the same direction as $F$. Overdamped models~\cite{asfaw04,asfaw-bekele2007,ai05+06}
predict  that for $F>F_{\rm stall}$  Brownian particles will
transfer heat against the temperature gradient,
from the cold to the hot bath, functioning as
a heat pump and at the stalled state the coefficient of
performance $\varepsilon$ will attain the Carnot coefficient of performance
$\varepsilon_{\rm c}$.

In Fig.~\ref{fig:fridge_nonuniform}, we plot the total power input ($\dot{W}$), various heat flows
out of the hot and cold cells as a function of the external load including
the entropy production and the coefficient of performance of the heat pump.
In contradiction to results obtained using overdamped
models~\cite{asfaw04,asfaw-bekele2007,ai05+06}, we observe that the stalled
state at which the current changes its direction is not the transition point 
from the heat engine to the heat pump. At the stalled state, the heat flow via potential,
which follows the direction of the particle flux $\langle v \rangle$,
goes to zero. However,  there is still a  net heat transfer from
the hot to the cold cell due to the kinetic
energy contribution. In fact, the heat flow via kinetic energy never vanishes
and always flows from the hot to the cold reservoir regardless
of the direction of particle current and magnitude of the external load.
The positivity of the entropy production at all values of the external force also
illustrates the validity of the second law  for our system.

\begin{figure}
\begin{center}
\leavevmode
\includegraphics[width=3.0in]{fig6.eps}
\caption{\label{fig:fridge_dt} (a) Power input $\dot{W}$ (blue double dash-dotted line), 
 various components of
$\dot{Q}_{2}$ [from top to bottom the heat flows are $\dot{Q}^\text{\sc
PE}_{2}$ (green line),
$\dot{Q}_{2}$ (orange line), $\dot{Q}^\text{\sc KE}_{2}$ (red line) and $\dot{Q}^\text{\sc J}_{2}$ (black line)] and
the entropy production $\dot{S}$ (black double dot-dashed line) and,
(b) the
coefficient of performance $\varepsilon$
as a function of $\Delta T$ for $M/m=5.0$, $\sigma_{B}=5.0$, $|F|L/2U_{0}=0.5$.
$\dot{W}$ and $\dot{Q}^\text{\sc PE}_{2}$ coincide because the external force $F$
is half of the force due to the potential i.e. $-2U_{O}/L$. Eqs.(~\ref{eq:work}) and ~(\ref{eq:heat_inertial_Langevin}) 
confirm this equality.
In (a), the thin horizontal
dashed line represents $y=0$. Temperature
of the cold bath is kept constant at $T_{2}=0.5$ and $T_{1}$ is varied.
$\varepsilon_{c}$ varies from
$\infty$ at $\Delta T=0$ to $20$ at $\Delta T =0.025$.}
\end{center}
\end{figure}

Figure~\ref{fig:fridge_nonuniform} shows that starting from
the stall load there is a  region
where the model works neither as a heat engine nor as a heat pump i.e.,
 heat flows from the hot to the cold cell while work is being
done on the Brownian particle. Refrigeration starts only
when the heat flow via potential exceeds the sum of joule heat
dissipated to the cold bath and the heat transfer via kinetic energy.
As the external force is further enhanced, the
joule heat (varying as $|F|^{2}$),
which originates from the power input  $\dot{W}$ and the frictional force in 
case of the heat pump, 
eventually dominates
the heat flow via potential (varying as $|F|$) and the cooling stops at
a certain large value of $F$. Beyond
this value of the external force, $\dot{Q}_{\rm 2}$ is negative showing
that net heat is also dissipated to the cold bath. The negative value
for the coefficient of performance correspond to this situation, where the
refrigeration fails.
There is an optimal load at which the heat flow out of
the cold cell and the coefficient of
performance attain a maximum but $\varepsilon$  is far below $\varepsilon_{c}$,
due to the irreversible heat flow via kinetic energy.

In Fig.~\ref{fig:fridge_dt}, we show the various
thermodynamic quantities as a function of the temperature difference between the cells. We
keep $T_{2}$
constant and vary $T_{1}$. As the temperature difference between the baths
increases,
the net negative current in the direction of $F$ will decrease as the increasing
temperature
of the hot bath will try to drive the Brownian particle in the forward
direction. This
decreases the heat flow via potential, which is proportional to the current,
and hence the heat pump's ability to transfer heat against the thermal
gradient. The joule heat also decreases with $\Delta T$. However, as $\Delta T \ll
1$, this decrease
is marginal. On the other hand, the irreversible heat
dumped to the cold cell via kinetic energy,
increases significantly with $\Delta T$
and greatly reduces the cooling. In fact, beyond a
small
temperature difference ${\Delta T} \ll 1$, refrigeration fails, as
Fig.~\ref{fig:fridge_dt} shows.
The coefficient of performance is monotonically reduced with increasing
temperature
difference between the cells and is much smaller compared to the Carnot
coefficient of performance due to the large entropy production.

\begin{figure}
\begin{center}
\leavevmode
\includegraphics[width=3.0in]{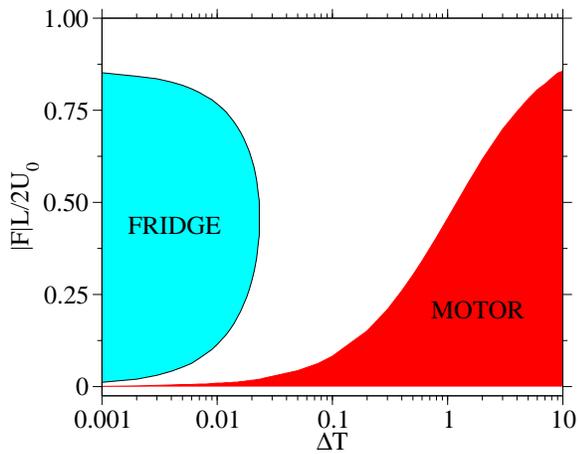}
\caption{\label{fig:fridge_motor}
The cyan and red areas represent the region where the system works
as a fridge (heat pump) and motor (heat engine) respectively. The white area represents the region
where the system acts neither as a heat engine nor as a heat pump. $M/m=5.0$,
$\sigma_{B}=5.0$. $\Delta T$ is varied by keeping initially $T_{1}=T_{2}=0.5$ and varying 
$T_{1}$.}
\end{center}
\end{figure}

In Fig.~\ref{fig:fridge_motor} we plot a phase diagram for our system in the $F-\Delta T$ plane.
Clearly, the motor (heat engine) and fridge (heat pump) regions do not touch 
each other (unlike overdamped models where the boundaries of the heat pump and heat engine 
regions touch~\cite{asfaw2013,asfaw-bekele2007}) 
and there is a large region (white space) where the system works 
neither as a heat engine nor as a heat pump. This corresponds to the situation
where a net heat is dissipated to the cold bath (due to the kinetic energy 
contribution or the joule heat) while the Brownian particles move in the
direction of the external force.

\section{Linear Irreversible Thermodynamics}
\label{sect:linear}
In this section we  discuss the efficiency and coefficient of performance of the
BL system in the linear response regime.
The BL heat engine and heat pump are a result of a cross effect between the
external force $F$ and the temperature difference ${\Delta T}=T_{1}-T_{2}$.  If
they are small, there exists a linear relationship between the thermodynamic
fluxes ($\langle v \rangle$ and
$\dot{Q}_{1 \rightarrow 2}$) and thermodynamic forces ($F/T$ and ${\Delta
T}/T^{2}$):
\begin{equation}
\label{eq:linear_response}
\begin{gathered}
\langle v \rangle  = L_{11}\frac{F}{T} + L_{12}\frac{\triangle T}{T^{2}},
\\
\dot{Q}_{1 \rightarrow 2}  = L_{21}\frac{F}{T} + L_{22}\frac{\triangle
T}{T^{2}},\\
\end{gathered}
\end{equation}
where $L_{ij}$'s ($i,j=1,2$) are the Onsager transport coefficients,
$T_{1}=T+{\Delta T}/2$
and $T_{2}=T-{\Delta T}/2$.

The efficiency of the heat engine is given by,
\begin{equation}
\label{eq:eff_linear}
\eta=\frac{-\dot{W}}{\dot{Q}_{1 \rightarrow 2}}= \frac{-F
\left(L_{11}\frac{F}{T} + L_{12}\frac{\triangle T}{T^{2}}\right)}
{L_{21}\frac{F}{T} + L_{22}\frac{\triangle T}{T^{2}}}\;.
\end{equation}
Recently, Van den Broeck~\cite{vandenbroeck07} investigated the
efficiency of Brownian heat engines
in the linear response regime and concluded that, in principle,
Carnot efficiency can be attained. He argued that for small $F$,
the temperature difference $\Delta T$ drives the heat engine in the
forward direction and simultaneously there is a heat transfer from the
hot to the cold reservoir. As  $F$ increases, we ultimately
reach the quasistatic condition of zero velocity
at  the stall load $F_{stall}$. From
Eq.~(\ref{eq:linear_response}), we find that the stall load is given by,
\begin{equation}
F_{stall}=-\frac{L_{12}}{L_{11}}\left(\frac{\Delta T}{T}\right)\;.
\end{equation}
Beyond the stall load, the Brownian particles reverse their direction, transferring heat
against the temperature gradient as a heat pump. In an ideal system, the
velocity
and heat reverse directions at the same value of $F_{stall}$ so that $\langle v
\rangle$
and $\dot{Q}_{1 \rightarrow 2}$ vanish for non-zero $F$ and $\Delta T$.
Substituting the value of $F_{stall}$ for $F$ in Eq.~(\ref{eq:eff_linear}), we
find that
this happens only when the condition of zero entropy production:
$L_{11}L_{22}=L_{12}L_{21}$
holds, and Onsager symmetry relationship is valid. This condition is referred to
as
``tight coupling" condition and when it is
satisfied, we find that efficiency at the quasistatic limit is equal to the
Carnot efficiency: ${\Delta T}/{T}$.

Carnot efficiency, however,  can only be reached in the
quasistatic condition where the velocity of the Brownian motor
is zero, and
hence it does not deliver any work. In practice a motor,
 must perform finite work in a finite
time. A more practical measure of the motor efficiency
is the efficiency when it delivers maximum power.
Curzon and Ahlborn~\cite{curzon75} claimed that the efficiency at
maximum power $\eta^{\ast}$ is bounded by the Curzon-Ahlborn efficiency,
$\eta^{\ast}_{max}=1-\sqrt{T_{2}/T_{1}}$, where $T_{1}>T_{2}$. The claim
was found to be valid for thermal motors~\cite{vandenbroeck05}.
Using overdamped models, Asfaw and Bekele~\cite{asfaw04} studied the efficiency at maximum power
($\eta^{\ast}$) of the BL heat engine  and found  that
$\eta^{\ast}$ approaches $\eta^{\ast}_{max}$ as the  temperature difference
between the
reservoirs $T_{1}-T_{2}=\Delta T \rightarrow 0$.
Apart from the fact that they used overdamped models, their
definition of work was different from the
the thermodynamic definition.
In order to evaluate efficiency at maximum power, we must use the
proper thermodynamic definition of work (Eq.~(\ref{eq:work})) and 
since heat flow via kinetic energy is the dominant source of heat
transfer for the BL system, one must also take into
account the inertia of the Brownian particle.

Maximum power will be attained at
a value of the load $F^{\ast}$ given by~
\cite{vandenbroeck05},
\begin{equation}
F^{\ast}=-\frac{L_{12}}{L_{11}}\left(\frac{\Delta T}{2T}\right)\;.
\end{equation}
Substituting the value of $F^{\ast}$ in Eq.~(\ref{eq:eff_linear}),
we find that the efficiency at maximum
power is given by,
\begin{equation}
\eta^{\ast}_{q}=\frac{\Delta T}{2T}\frac{q^{2}}{2-q^{2}}\;,
\end{equation}
where,
\begin{equation}
q=\frac{L_{12}}{\sqrt{L_{11}L_{22}}}\;.
\end{equation}
If the necessary condition of zero entropy production is satisfied,
then $q=1$ and  to the lowest order of $\Delta T$,
the efficiency at maximum power approaches
the efficiency of an endoreversible heat engine at maximum power viz., the
Curzon-Ahlborn
efficiency: $\eta^{\ast}_{max}=1-\sqrt{T_{2}/T_{1}}\sim {\Delta
T}/{2T}$~\cite{curzon75}.
However, in a
previous work~\cite{benjamin08} it was shown that the product of the diagonal
coefficients  $L_{11}L_{22}$ diverges as $(M/m)^{-1/2}$, reflecting the
divergence of the kinetic energy contribution. The off-diagonal coefficients
however, did not show this singular behaviour. Consequently, we find
$L_{11}L_{22}\gg L_{12}L_{21}$ for
all values of $M/m$ indicating large entropy production and very low efficiency.
Hence the BL heat engine can never attain the Carnot efficiency
or the Curzon-Ahlborn efficiency.

The coefficient of performance of the heat pump is given by,
\begin{equation}
\label{eq:lin_cop}
\varepsilon=\frac{-\dot{Q}_{1 \rightarrow 2}}{\dot{W}} =
\frac{-L_{21}\frac{F}{T}-L_{22}\frac{\Delta T}{T^{2}}}
{F\left(L_{11}\frac{F}{T}+L_{12}\frac{\Delta T}{T^{2}}\right)}
\end{equation}
By similar arguments as in the case of the heat engine, it is easy to show that the
coefficient of performance  reaches the
Carnot coefficient of performance $T/{\Delta T}$ in the
quasistatic limit, only if the necessary
condition $L_{11}L_{22}=L_{12}L_{21}$ holds.

In the quasistatic limit, the heat pump does not transfer any heat and hence
is practically useless. Since Carnot coefficient of performance is only possible
in the quasistatic limit, it is not an effective measure of the
heat pump performance from a practical point of view. A heat pump
must transfer  heat in a finite time and
similar to the efficiency
at maximum power for the motor we must find the coefficient of
performance of the heat pump, under conditions of maximum 
heat transfer from the cold to the hot bath.
Following Van den Broeck's work~\cite{vandenbroeck05},
regarding the efficiency at maximum power for the motor,  some authors have
investigated
the coefficient of performance of
the refrigerator
at maximum $\dot{Q}_{1 \rightarrow 2}({\Delta T}/T)$~\cite{cisneros06}.
This quantity is maximum at a temperature difference ${\Delta T}^{\ast}$, given
by
\begin{equation}
\label{eq:DT_maxQ}
{\Delta T}^{\ast}=-\frac{L_{21}}{L_{22}}\frac{FT}{2}\;.
\end{equation}
Substituting the value of ${\Delta T}^{\ast}$ in place of $\Delta T$
in Eq.(~\ref{eq:lin_cop}),
we find that up to the lowest order
of $\Delta T$, the coefficient of performance at maximum
$\dot{Q}_{1 \rightarrow 2}({\Delta T}/T)$, is given by
\begin{equation}
\label{eq:cop_maxQ_DT}
\varepsilon^{\ast}_{q}=\frac{q^{2}}{2-q^{2}}\frac{T}{2{\Delta T}}=\frac{L_{21}/L_{11}}{\mid F\mid(2-q^{2})}
\end{equation}
Analogous to the efficiency at maximum power, $\varepsilon^{\ast}$
is equal to a factor that depends on $q$ only, times half the
Carnot coefficient of performance.
If the necessary condition for attaining the Carnot
coefficient of performance, namely $L_{11}L_{22}=L_{12}L_{21}$ holds, then
$q=1$  and  $\varepsilon^{\ast}$ attains a maximum $\varepsilon^{\ast}_{max}$
given by,
\begin{equation}
\label{eq:cop_maximum}
\varepsilon^{\ast}_{max}=\frac{T}{2{\Delta T}}=\frac{L_{21}}{L_{11}}\frac{1}{2|F|}
\end{equation}

However, for the BL refrigerator, $q=1$ will never be attainable
due to the kinetic energy contribution~\cite{benjamin08} and consequently
$\varepsilon^{\ast}_{q}$ will be less than $\varepsilon^{\ast}_{max}$.

These predictions of linear irreversible thermodynamics concerning
$\eta^{\ast}$ and $\varepsilon^{\ast}$ have not been verified when temperature
is spatially inhomogeneous. One of our objectives is to investigate the
validity of linear irreversible finite time thermodynamics for the
BL heat-engine and heat pump via numerical simulation of the
inertial Langevin equation.

\begin{figure}
\begin{center}
\leavevmode
\includegraphics[width=3.0in]{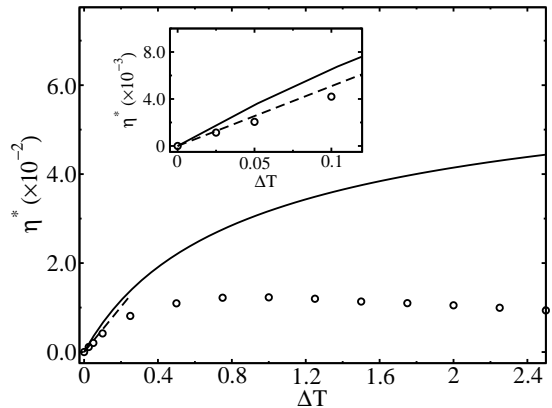}
\caption{\label{fig:motor_effmaxpowermassDT}  $\eta^{\ast}$ (symbols),
$\eta^{\ast}_{q}$ (dashed line) and the
Curzon-Ahlborn efficiency $\eta^{\ast}_{max}$ (solid line) as a function of
${\Delta T}=T_{1}-T_{2}$, where the temperature
of the cold cell is kept at $T_{2}=0.5$. $\eta^{\ast}_{max}$ has been scaled
($\eta^{\ast}_{max}\times 0.075$)
for easy visualization of all quantities on the same plot. The inset shows the
detail of the
efficiency at maximum power in the linear response regime. Parameter values are
$M/m=5.0$ and $\sigma_{B}=0.5$.}
\end{center}
\end{figure}

From Fig.~\ref{fig:motor_qke} we find that the power output $-\dot{W}$ reaches a
maximum
at an optimum value of the  load $F^{\ast}$. We evaluates the efficiency
at $F^{\ast}$, which is the efficiency at maximum power, as a function of
$\Delta T$.
In Fig.~\ref{fig:motor_effmaxpowermassDT}, we plot as a function of
$\Delta T$, the Curzon-Ahlborn
efficiency $\eta^{*}_{max}$, $\eta^{*}$ obtained from numerical solution of
the inertial Langevin equation and the efficiency at maximum power
predicted by linear response theory, namely $\eta^{\ast}_{q}$. In order
to determine $\eta^{\ast}_{q}$ we have to find $q$, which was evaluated
from the values of the Onsager coefficients, numerically. 
As the kinetic energy contribution
is proportional to the temperature difference, it  diminishes
$\eta^{*}$ to zero as $\Delta T\rightarrow \infty$, thus contradicting the
results
of Ref.~\cite{asfaw04}, obtained using overdamped models. There is an
optimum $\Delta T$ at which $\eta^{*}$ reaches a weak maximum but it is
far below $\eta^{*}_{max}$. In the linear response regime, there is
good agreement between $\eta^{\ast}$ and  $\eta^{\ast}_{q}$, as shown in the
inset
of Fig.~\ref{fig:motor_effmaxpowermassDT}.

\begin{figure}
\begin{center}
\leavevmode
\includegraphics[width=3.0in]{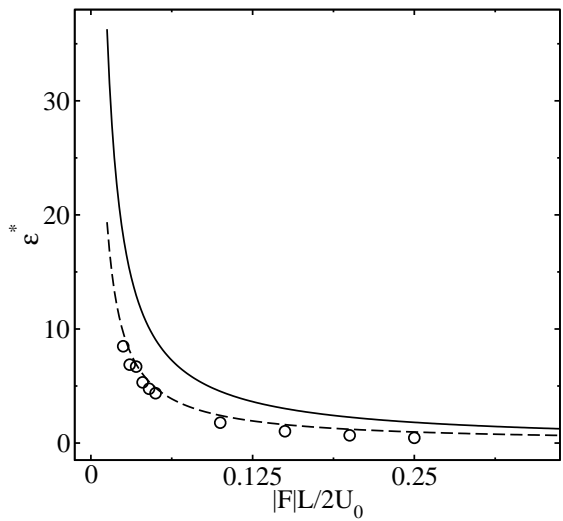}
\caption{\label{fig:linear_cop_vs_f} $\varepsilon^{\ast}$ obtained
from numerical simulation (symbols), $\varepsilon^{\ast}_{q}$ (dashed line) and
$\varepsilon^{\ast}_{max}$ (solid line), as a function of the external force.
Parameter values are $M/m=5.0$, $\sigma_{B}=5.0$ and $T_{1}=T_{2}=0.5$.}
\end{center}
\end{figure}

Performance of the refrigerator at maximum $\dot{Q}_{2
\rightarrow 1}{\Delta T}/T$ was also investigated
numerically.
The quantity  $\dot{Q}_{2 \rightarrow 1}{\Delta T}/T$ reaches a maximum
at an optimum value of the temperature difference, namely ${\Delta T}^{\ast}$.
We evaluated the coefficient of performance $\varepsilon^{\ast}$
at ${\Delta T}^{\ast}$ as a function of  $|F|$.
Figure~\ref{fig:linear_cop_vs_f} shows that $\varepsilon^{\ast}$ is always
lower than $\varepsilon^{\ast}_{max}$ due to the kinetic energy contribution.
There
is also good agreement between $\varepsilon^{\ast}$ and the coefficient of
performance
predicted by linear response theory, $\varepsilon^{\ast}_{q}$.

\section{Efficiency Optimization}
\label{sect:effopt}

Figure~\ref{fig:motor_mass}  showed
that $\eta$ ($\varepsilon$) reaches a maximum at  optimum value of the mass  between
the overdamped and underdamped limits. In this regime, the efficiency can
be further enhanced by reducing the kinetic energy contribution.

The main cause of the low efficiency and coefficient of performance
in the BL heat engine and heat pump respectively is the heat transfer
due to kinetic energy across the temperature boundary. When a temperature
boundary coincides with the potential minima, the situation is the worst
since the Brownian particle crosses the temperature boundary many times,
resulting in a large heat transfer via kinetic energy. In order to reduce the number of crossings,
we must move the temperature boundary away from the potential minima. On the
other hand, the other temperature boundary should remain close to the
potential maxima to maintain the current of Brownian particles.

Another way to minimize the number of crossings would be to keep
the same temperature profile but modify the potential shape so that there
is a small potential barrier of height $E<U_{0}$
at the location of the potential minima, which in our original model
coincides with the temperature boundary. Since the
Brownian particle is less likely to be found at a potential maxima,
the number of crossings will decrease and consequently the
kinetic energy contribution will be diminished.

Based on the above ideas, we will look for better models
for the BL heat pump and heat engine by considering two approaches, one
where we vary the temperature profile  ({\textit{case $I$}})  and the other
where the potential shape ({\textit{case $II$}}) is varied as compared to our original model.

While the efficiency reaches a maximum in between the overdamped and
underdamped regimes, most microscopic systems can be described to a good
approximation using overdamped dynamics. Moreover, most studies on the efficiency
optimization of the BL heat engine and heat pump have been carried out in the
overdamped regime using overdamped models~\cite{asfaw04,asfaw2013,asfaw-bekele2007,asfawepjb2008,ai05+06,matsuo00}.
However, as the overdamped model fails to predict the correct heat transfer we use
the inertial Langevin equation [Eq.~(\ref{eq:Langevin})] to study the efficiency
optimization in the overdamped regime by choosing a large value for the friction
coefficient, where good agreement is observed between the inertial Langevin equation
and the overdamped model~\cite{benjamin08}. To this end, we carry out  numerical simulations 
of Eq.~(\ref{eq:Langevin}) in the overdamped regime ($\gamma_{\rm 1,2}> 10.0$) with a 
dimensionless timestep $\tau=0.0001$.

\subsection{Heat Engine}
\label{subsect:effopt_motor}

\begin{figure}
\begin{center}
\includegraphics[width=3.0in]{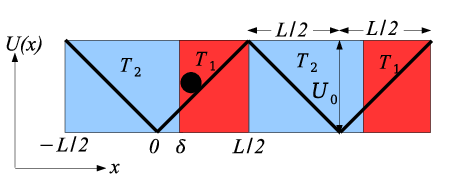}
\caption{\label{fig:modelmot_1}(Color online)  The temperature and
potential profiles corresponding to {\textit{case $I$}} of heat engine and
heat pump. 
The dark circle represents the Brownian particle.}
\end{center}
\end{figure}

In  {\textit{case I}},  we  shift the location of the temperature
boundary away from the potential minima. The
potential  $U(x)$ is the same as in our original model but  the temperature
profile $T(x)$ is modified and  is now given by:
\begin{equation}
\label{eq:T_a}
T(x) =
\begin{cases}
T_{2} & \text{for $-\frac{\displaystyle L}{\displaystyle 2} < x \leqslant
\delta$}
\\
T_{1} & \text{for $\delta < x \leqslant L$}\\
\end{cases}
\end{equation}
where, $T_{1}>T_{2}$ and the parameter $\delta$ specifies the location
of the temperature boundary.

\begin{figure}
\begin{center}
\leavevmode
\includegraphics[width=3.0in]{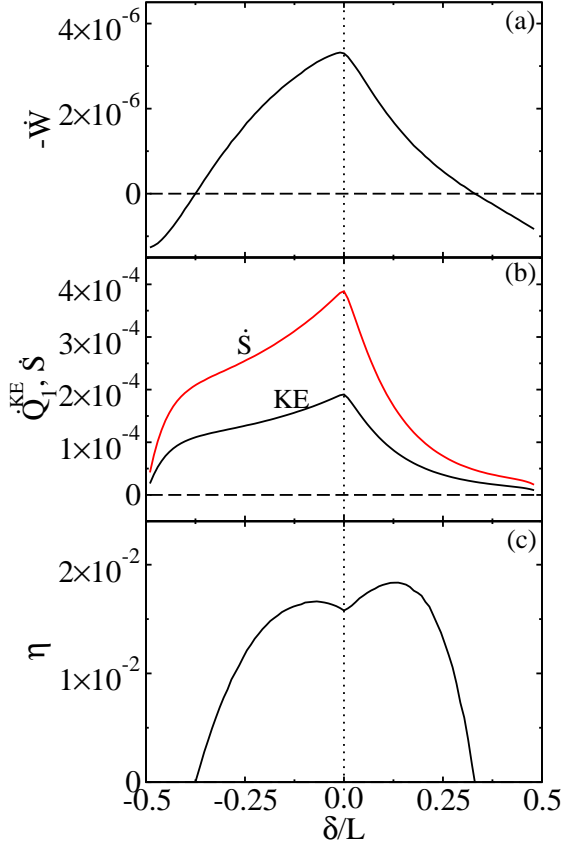}
\caption{\label{fig:motor_caseI} (a) Power output ($-\dot{W}$), (b) entropy production
and heat flux due to
kinetic energy, and (c) efficiency  as a function of $\delta/L$
for {\textit{case I}} of heat engine,
obtained from the inertial Langevin equation. In (a) and (b), the thin
horizontal dashed line represents $y=0$. Parameter values are $T_{1}=0.7$,
$T_{2}=0.3$, $|F|L/2U_{0}=0.1$, $M/m=5$ and $\sigma_{B}=5.0$. The
Carnot efficiency $\eta_{c}=0.57$.}
\end{center}
\end{figure}

At $\delta/L=0$,
the cell boundary coincides with the potential maxima/minima and we go back
to our previous model studied in section~\ref{sect:basic_model}. At
$\delta/L=-1/2$
and $1/2$,
the temperature becomes uniform since the width of the cold
(${\delta}/L=-1/2$)/hot
(${\delta}/L=1/2$) cell goes to zero and
the system ceases to work as a heat engine. The heat engine can deliver power for $\delta/L$
in the range
$-1/2<{\delta}/L<1/2$. In Fig.~\ref{fig:motor_caseI},
 we plot the power output, heat flow and entropy production and
efficiency as a function of the parameter $\delta/L$.
Figure~\ref{fig:motor_caseI}
clearly shows that
as the cell
boundary moves away from the potential minima,  in either direction,
the irreversible heat transfer via
kinetic energy and the entropy production decrease.

When $\delta/L>0$, the presence
of the cold region reduces the likelihood of the Brownian particle
jumping over the barrier in the positive direction, especially if
$(2U_{0}/L)\delta\gg T_{2}$,
thus diminishing the current and the power $-\dot{W}$, and consequently
the efficiency. On the other hand if $\delta/L$ is close to the potential
minima,
the kinetic energy contribution becomes large, again reducing the efficiency.
At an optimum $\delta/L>0$, the efficiency is slightly maximized as shown in
Fig.~\ref{fig:motor_caseI}.

When $\delta/L<0$, the hot region extends to the left of the potential minima at
$x=0$.
This enhances the probability of the Brownian particle
reaching the higher-potential-energy region and crossing the barrier in the
negative
direction. As a result, the net particle current in the positive
direction and the power delivered by the Brownian
particle diminish, thus reducing the efficiency. On the
other hand, if $\delta/L$
is close to $x=0$, the large heat transfer via kinetic energy
reduces the efficiency. Figure~\ref{fig:motor_caseI} shows that
there is an optimum $\delta/L<0$, at which the efficiency is maximized. However,
the maximum efficiency is far below the Carnot limit.

\begin{figure}
\begin{center}
\includegraphics[width=3.0in]{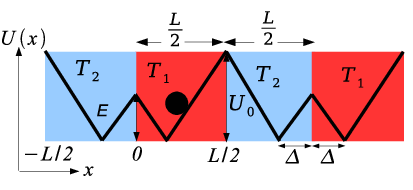}
\caption{\label{fig:modelmot_3}(Color online)  The temperature and
potential profiles corresponding to {\textit{case $II$}} of heat engine and
heat pump.
The dark circle represents the Brownian particle.}
\end{center}
\end{figure}

\begin{figure}
\begin{center}
\leavevmode
\includegraphics[width=3.0in]{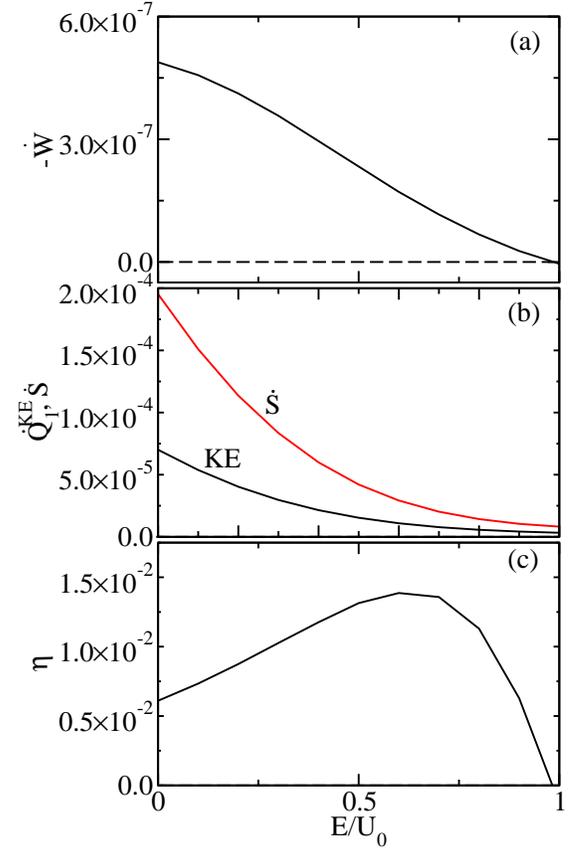}
\caption{\label{fig:motor_caseII} (a) Power output, (b) entropy production and heat
flux due to kinetic energy, and (c) efficiency  as a function of
$E/U_{0}$ for {\textit{case II}} of heat engine. In (a), 
the thin horizontal
dashed line represents $y=0$. Parameter values are $T_{1}=0.4$, $T_{2}=0.2$,
$\Delta/L=0.1$, $|F|L/2U_{0}=0.025$, $M/m=5$ and $\sigma_{B}=5.0$. The
Carnot efficiency $\eta_{c}=0.5$.}
\end{center}
\end{figure}

Now, we come to the second model.
In {\textit{case II}}, we keep the
same temperature profile as in our original model [Eq.~(\ref{eq:T})]
but change the potential profile $U(x)$, which  is now given by,
\begin{equation}
\label{eq:pot_b}
U(x) =
\begin{cases}
-\frac{2U_{0}}{L-2\Delta} (x+\Delta) & \text{for $-\frac{\displaystyle
L}{\displaystyle 2} < x \leqslant -\Delta$} \\
\frac{E}{\Delta} (x+\Delta) & \text{for -$\Delta < x \leqslant 0 $}\\
-\frac{E}{\Delta} (x-\Delta) & \text{for $0 < x \leqslant \Delta $}\\
\frac{2U_{0}}{L-2\Delta} (x-\Delta) & \text{for $\Delta < x \leqslant
\frac{\displaystyle L}{\displaystyle 2}$}
\end{cases}
\end{equation}
Equation(~\ref{eq:pot_b}) shows that the potential has a  barrier of
height $E<U_{0}$ at $x=0$, coinciding with the position of the potential minima
in our
original model.

In Fig.~\ref{fig:motor_caseII}, we plot $-\dot{W}$, various components
of $\dot{Q}_{1}$ and $\eta$ as a function of the barrier height $E$ at
$x=0$ normalized by $U_{0}$.
At $E/U_{0}=0$, the temperature boundary coincides with
a potential minimum and the kinetic energy contribution is large thus
leading to a low efficiency. As we increase $E/U_{0}$,
the irreversible heat transfer via kinetic
energy as well as the net entropy production decreases as the recrossing
of the Brownian particle near the temperature boundary at $x=0$ is
reduced on account of a potential maximum there. However, the motor velocity and
consequently the power  is diminished as well because the asymmetry around a
potential minimum due to the inhomogeneous temperature decreases. The
efficiency is maximized for $0<E/U_{0}<1$, as shown in
Fig.~\ref{fig:motor_caseII},
though it remains far
below the Carnot limit ($\eta_{C}=0.5$).

\subsection{Heat Pump}
\label{sect:effopt_fridge}

\begin{figure}
\begin{center}
\leavevmode
\includegraphics[width=3.0in]{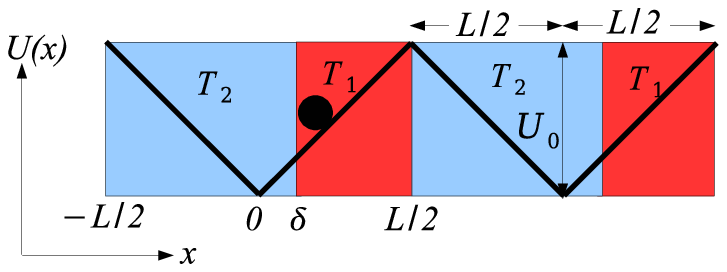}
\caption{\label{fig:fridge_model1} (a) Power input $\dot{W}$, 
various components of $\dot{Q}_{2}$ (from top to bottom, the heat flows are
$\dot{Q}^\text{\sc PE}_2$, $\dot{Q}_{2}$, $\dot{Q}^\text{\sc J}_2$ and
$\dot{Q}^\text{\sc KE}_2$) and entropy production and, (b) the coefficient of performance $\varepsilon$
as a function of $\delta/L$, corresponding to {\textit{case
I}} of heat pump,  obtained from numerical solution of 
the inertial Langevin equation. In (a),  the thin horizontal
dashed line represents $y=0$. The parameter values are
$|F|L/2U_{0}=0.2$, $M/m=5.0$ and $\sigma_{B}=5.0$, $T_{1}=0.51$ and $T_{2}=0.5$.
The Carnot coefficient of performance $\varepsilon_{c}=50$.}
\end{center}
\end{figure}

We now consider the performance of the BL heat pump
corresponding to the same two cases, which were introduced in the previous
subsection for
the BL heat engine.

In Fig.~\ref{fig:fridge_model1}, we plot the power input, heat flow out of
cell 2, the entropy production and the coefficient of performance as a function of the location
of the temperature boundary $\delta/L$ corresponding to {\textit{case I}.
Since ${\Delta T}\ll 1$, the thermal gradient has a negligible effect on the
particle current. Instead,
the magnitude of the negative current is determined mostly by the potential
shape
and the external force $F$. However, the potential profile stays the same
regardless
of the position of the temperature boundary. Hence, the power input varies
slowly
with $\delta/L$, as shown in Fig.~\ref{fig:fridge_model1}.

As expected, Fig.~\ref{fig:fridge_model1} shows that
the kinetic energy contribution and the entropy production become smaller as the temperature boundary is
shifted away from the potential minima. The joule heat dissipated to the cold
cell vanishes at $\delta/L=-1/2$ since the width of the cold region goes to zero
at that point. As $\delta/L$ shifts to the right of $x=-1/2$, the joule heat
starts to increase as the width of the cold cell becomes larger and reaches a
maximum
at $\delta/L=1/2$, when the cold cell occupies the entire period $L$ of the
potential.
As the temperature boundary is shifted to the left of $\delta/L=0$,
the heat flow via potential  $\dot{Q}_{2}^{\text{PE}}$,
is reduced since
the width of the cold cell becomes smaller  and
less potential energy is extracted.
When $\delta/L>0$,
$\dot{Q}_{2}^{\text{PE}}$  again diminishes
since the cold cell now includes a portion of the negative slope as well.
Figure~\ref{fig:fridge_model1} shows that
there is an optimum  $\delta/L<0$ at which the total
heat extracted from cell 2 and the coefficient of performance  reach
a maximum, showing that this model can enhance the cooling and
refrigerator performance by appropriately changing the
location of the temperature boundary. However, $\varepsilon$ is
still far below the Carnot coefficient of performance.

In {\textit{ case II}}, the small barrier at $x=0$ reduces the heat transfer
via kinetic energy and consequently the entropy. This is clearly seen in  Fig.~\ref{fig:fridge_model2},
 Due to the
small temperature difference, the current and consequently the power input
and the joule heat vary slowly with $E/U_{0}$. On the other hand,
the barrier of height $E$ at $x=0$ diminishes the heat flow via potential
because due to
the negative slope, $-E/\Delta$, an amount of heat
$E$ is dissipated each time the Brownian particle crosses the cold cell, thus
reducing the magnitude of potential energy extracted to  $U_{0}-E$.
Since, heat flow via potential decreases with $E/U_0$ more rapidly as compared
to the kinetic energy contribution, there is no optimum value of  $E/U_{0}$
and net heat extracted from cell 2 decreases monotonically from a maximum at
$E/U_{0}=0$. The coefficient
of performance also decreases with $E/U_{0}$ and
is far below $\varepsilon_{c}$. Hence, this case does not improve
the performance of the refrigerator.

\begin{figure}
\begin{center}
\leavevmode
\includegraphics[width=3.0in]{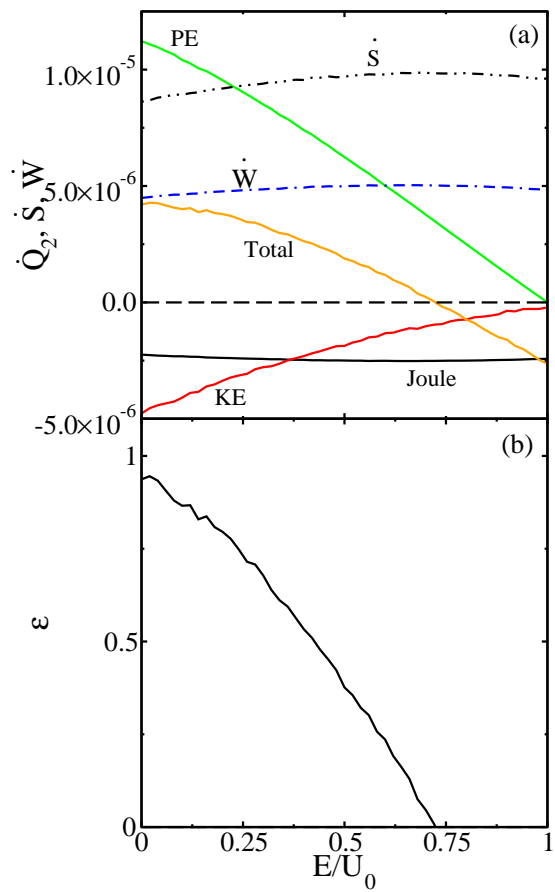}
\caption{\label{fig:fridge_model2}(a) Power input $\dot{W}$,
various components of $\dot{Q}_{2}$ (from top to bottom, the heat flows are
$\dot{Q}^\text{\sc PE}_2$, $\dot{Q}_{2}$, $\dot{Q}^\text{\sc J}_2$ and
$\dot{Q}^\text{\sc KE}_2$) and entropy production and, (b) the coefficient of performance $\varepsilon$
as a function of $E/U_{0}$, corresponding to {\textit{case
II}} of heat pump, obtained from
the inertial Langevin equation. In (a), the thin horizontal
dashed line represents $y=0$. The parameter values are
$|F|L/2U_{0}=0.2$, $M/m=5.0$ and $\sigma_{B}=5.0$, $T_{1}=0.51$ and $T_{2}=0.5$.
The Carnot coefficient of performance $\varepsilon_{c}=50$. }
\end{center}
\end{figure}

\section{Conclusion}
We studied the efficiency and coefficient of performance of the
Buttiker-Landauer motor and  refrigerator by
numerical simulation of the inertial Langevin equation.
Our results show qualitatively different behaviour 
of the heat transfer, efficiency and coefficient of performance
as a function of various parameters as compared to previous works based on
overdamped models
or other phenomenological approaches.

There is
an optimal value of the mass  at which the
efficiency reaches a maximum. However, it is far below the
Carnot limits. While there exists an optimal temperature difference between the
hot and cold baths at which the efficiency of the motor is maximized the refrigerator
performs best only when the temperature difference is small.
We also found that
the efficiency at maximum power can never attain the Curzon-Ahlborn efficiency
due to the irreversible heat flow via kinetic energy. Due to this irreversible
heat,
the coefficient of performance of the refrigerator is below the theoretical
maximum
predicted by linear response theory, under conditions which could
be considered as the equivalent of maximum power. The predictions of linear
irreversible thermodynamics are in good agreement with numerical data.
For the models we have studied,
changing potential shape or the temperature profile
in order to reduce the heat flow via kinetic energy
reduces the particle current as well, leading only to a small
enhancement of the motor efficiency and refrigerator
coefficient of performance. We have performed numerical simulations of a few other models
introduced by other authors in the context of efficiency optimization but in all cases
decrease of kinetic energy also reduced the power output and the gain in efficiency
was negligible.

From our results it is observed that reduction of heat transfer via kinetic energy 
also leads to a diminishing of the
particle current thus making the motor marginally more
efficient. In a recent work, Berger {\it et al.}~\cite{berger2009} followed 
another approach where they optimized the potential to maximize the power 
output for a piecewise linear temperature profile. In fact the optimal potential obtained by 
them led to a diverging particle current in the overdamped limit. However, it is not clear
whether such a potential will also lead to a reduction of the irreversible heat
via kinetic energy and thus to a higher efficiency.

The  next question to address
is how one can reduce the
irreversible heat without decreasing the particle current. At a more fundamental
level,
it would be interesting to obtain theoretically the maximum efficiency at
maximum power
and the equivalent maximum  for the coefficient of performance. Answering such
questions
would also enhance our understanding of non-equilibrium thermodynamics at the
small scale.

\end{document}